\documentclass[12pt,usenames,dvipsnames]{article}

\usepackage{latexsym}
\usepackage{amssymb,amsfonts,amsmath}
\usepackage{graphicx} 
\usepackage{indentfirst}
\usepackage{bbm}
\usepackage{amssymb}
\usepackage{verbatim}
\usepackage{amsmath, amsthm,amssymb}
\usepackage{mathrsfs}
\usepackage{hyperref}
\usepackage{amsfonts}
\usepackage{dsfont}
\usepackage{cite}
\usepackage{xcolor}
\usepackage{enumerate}
\usepackage{cleveref}

\parskip=\medskipamount

\newcommand {\cD}{{\cal D}}

\newcommand {\cL}{{\cal L}}

\newcommand {\cN}{{\cal N}}
\newcommand {\cO}{{\cal O}}

\def\a{\alpha}

\def\b{\beta}

\def\d{\delta}

\def\g{\gamma}

\def\l{\lambda}

\def\t{\tau}

\def\O{\Omega}

\def\rd{{\rm d}}
\def\ri{{\rm i}}
\def\re{{\rm e}}

\newcommand{\ve}{\varepsilon}                            %

\newcommand{\pa}{\partial}                           %
\newcommand{\hf}{\frac12}

\newcommand{\vf}{\varphi}

\newcommand{\be}{\begin{equation}}
\newcommand{\ee}{\end{equation}}
\newcommand{\bea}{\begin{eqnarray}}
\newcommand{\eea}{\end{eqnarray}}
\newcommand{\non}{\nonumber}

\def\double #1{#1{\hbox{\kern-2pt $#1$}}}

\newif\ifdtup

\newcommand{\bsubeq}{\begin{subequations}}
\newcommand{\esubeq}{\end{subequations}}

\numberwithin{equation}{section}

\newcommand{\sSp}{\mathsf{Sp}}

\newcommand{\sSL}{\mathsf{SL}}

\newcommand{\sSO}{\mathsf{SO}}
\newcommand{\sO}{\mathsf{O}}
\newcommand{\sU}{\mathsf{U}}

\begin{document}

\begin{titlepage}
\begin{flushright}
January, 2026 \\
Revised version: June, 2026
\end{flushright}
\vspace{5mm}

\begin{center}
{\Large \bf On nonlinear self-duality in $4p$ dimensions}
\end{center}

\begin{center}

{\bf Sergei M. Kuzenko} \\
\vspace{5mm}

\footnotesize{
{\it Department of Physics M013, The University of Western Australia\\
35 Stirling Highway, Perth W.A. 6009, Australia}}  
~\\

\vspace{2mm}

\end{center}

\begin{abstract}
\baselineskip=14pt
Building on the earlier work by Araki and Tanii, Aschieri {\it et al.}, and Buratti {\it et al.}, we demonstrate that every model for self-dual nonlinear electrodynamics in four dimensions has a $\mathsf{U}(1)$ duality-invariant extension to $4p>4$ dimensions and construct new self-dual nonlinear theories for a gauge $(2p-1)$-form. We present a family of models for self-dual nonlinear $(2p-1)$-form electrodynamics in which the trace of the energy-momentum tensor determines the flow with respect to a duality-invariant deformation parameter. Finally, we propose the interesting problem of computing the so-called induced action in the case that self-dual $(2p-1)$-form electrodynamics is coupled to dilaton and axion fields and the compact duality group $\mathsf{U}(1)$ is enhanced to the non-compact group $\mathsf{SL}(2, {\mathbb R})$.
\end{abstract}
\vspace{5mm}

\vfill

\vfill
\end{titlepage}

\newpage
\renewcommand{\thefootnote}{\arabic{footnote}}
\setcounter{footnote}{0}

%%
%\tableofcontents{}
%\vspace{1cm}
%\bigskip\hrule

\allowdisplaybreaks

%%%%%%%%%%%%%%%%%%%%%%%%%%%%%%%%
%%%%%%%%%%%%%%%%%%%%%%%%%%%%%%%%

\section{Introduction}

As an extension of the seminal work by Gaillard and Zumino \cite{GZ1}, 
the general formalism of duality-invariant
models for nonlinear electrodynamics in four dimensions was developed in the mid-1990s \cite{GR1,GR2,GZ2,GZ3}, 
see \cite{KT2,AFZ} for comprehensive reviews.  
The Gaillard-Zumino-Gibbons-Rasheed  
(GZGR) approach was generalised to off-shell $\cN=1$ and $\cN=2$ globally \cite{KT1,KT2} and locally \cite{KMcC,K12} supersymmetric theories. In particular, the first consistent 
perturbative scheme to construct the $\cN=2$ supersymmetric Born-Infeld action 
was given in \cite{KT2} (this approach was further pursued in \cite{BCFKR, Ivanov:2013maa}). 
The GZGR formalism was also extended to higher dimensions
\cite{GR1, Tanii,AT,ABMZ}.

In Minkowski space of even dimension $d=4p \equiv 2n$, with $p$ a positive integer,
we  consider
a self-interacting theory of a  gauge $(n-1)$-form $A_{a_1 \dots a_{n-1}}$
with the property that its Lagrangian, $L (F_{a_1 \dots a_n})$,
 is a function of the field strengths 
$F_{a_1 \dots a_n} =n  \pa_{ [a_1} A_{a_2 \dots a_{n} ] } \equiv F_{a(n)}$.\footnote{We follow the notation and conventions of \cite{KT2}.}
We assume that the theory possesses $\sU(1)$ duality invariance. 
This means that the Lagrangian is a solution
to the self-duality equation \cite{AT}
\bea
\widetilde G^{a_1 \dots a_n} G_{a_1 \dots a_n} 
+ \widetilde F^{a_1 \dots a_n} F_{a_1 \dots a_n} \equiv \widetilde G \cdot G + \widetilde F \cdot F
=0~,
\label{self-duality}
\eea
where we have introduced
\bea
\widetilde{G}^{a_1 \dots a_n} (F)= n!
\frac{\pa L (F)}{\pa F_{ a_1 \dots a_n}}~.
\eea
As usual, the notation $\widetilde F$ is used  for the Hodge dual of $F$, 
\bea
\widetilde{F}^{a_1 \dots a_n} =
\frac{1}{n!} \, \ve^{a_1 \dots a_n b_1 \dots b_{n}} \,
F_{b_1 \dots b_{n} }~.
\eea
The simplest solution of \eqref{self-duality} is the free model
\bea
L_{\rm free}(F) = -\frac{1}{2 n!} F^{a_1 \dots a_n}  F_{a_1 \dots a_n} ~,
\label{free}
\eea
which is a conformal field theory in $d=2n$ dimensions. 

Only a few explicit solutions of the self-duality equation \eqref{self-duality} are known beyond four dimensions.
In particular,  Ref. \cite{ABMZ} introduced Born-Infeld-like actions with $N\geq 1$ complex field strengths (and also with $N>1$ real field strengths) as a generalisation of the Ro\v{c}ek-Tseytlin construction \cite{RT} for the Maxwell-Goldstone multiplet model \cite{BG} for partial 
 $\cN=2 \to \cN=1$ supersymmetry breaking \cite{Antoniadis:1995vb}.
In this letter we will demonstrate, building on the results of \cite{AT, ABMZ, Buratti:2019cbm},
 that every model for self-dual nonlinear electrodynamics in four dimensions has a $\sU(1)$ 
duality-invariant extension to $4p =2n >4$  dimensions and construct new self-dual nonlinear theories for a gauge $(n-1)$-form. We also present a family of models for self-dual $(n-1)$-form electrodynamics in which the trace of the energy-momentum tensor determines the flow with respect to a duality-invariant deformation parameter. 
Finally, we put forward an interesting open problem of computing a so-called induced action in the case that self-dual $(n-1)$-form electrodynamics is coupled to dilaton and axion fields and the compact duality group $\mathsf{U}(1)$ is enhanced to the non-compact 
group $\mathsf{SL}(2, {\mathbb R})$.

%%%%%%%%%%%%%%%%%%%%%%%%%%%%%%%%%%%%
%%%%%%%%%%%%%%%%%%%%%%%%%%%%%%%%%%%%

\section{Different forms of the self-duality equation}

It is useful to introduce (anti) self-dual components of $F$, 
\bea
F^\pm_{a_1\dots a_n} = \hf \Big( F_{a_1\dots a_n}  
\pm \ri \widetilde F_{a_1\dots a_n} \Big) ~, \qquad 
\widetilde F^\pm = \mp\ri F^\pm ~,
\eea
such that
\bea
F = F^+ +F^-~,\qquad \widetilde{F} = - \ri F^+ + \ri F^-~.
\eea
The same definition can be used to introduce the (anti) self-dual components of $G$. Then the self-duality equation 
\eqref{self-duality} turns into 
\bea 
{\rm Im} \big( G^+ \cdot G^+ + F^+ \cdot F^+\big) =0~.
\label{self-duality2}
\eea

We are going to look for a class of $\sU(1)$ duality-invariant theories of the form\footnote{In the $d =2n = 4p+2$ case, the invariant $P$ is identically equal to zero.} 
\bea
L(F_{a(n)}) = \cL (S, P)~, \qquad 
S= - \frac{1}{2n!} F^{a_1\dots a_n} F_{a_1\dots a_n} 
~, \qquad  P= - \frac{1}{2n!} \widetilde{F}^{a_1\dots a_n} F_{a_1\dots a_n}~.
\label{models}
\eea
In four dimensions, $S$ and $P$ are the only independent invariants of the electromagnetic field. More general invariants of the $n$-form  $F_{a(n)}$ in $d=2n=4p>4$ dimensions exist (see, e.g.,  \cite{Cederwall:2025ywy} for the technical details). However, we restrict our attention to models of the form \eqref{models}.
For the theories under consideration, the self-duality equation \eqref{self-duality2} is equivalent to 
\bea
P \big( \cL_S^2 - \cL_P^2 - 1 \big) = S\cL_S \cL_P~,
\label{self-duality3}
\eea
where $\cL_S$ and $\cL_P$ denote the derivatives of $\cL$ with respect to $S$ and $P$, respectively.
Eq. \eqref{self-duality3} proves to be the self-duality equation for nonlinear electrodynamics $(n=2)$ written in the above form for the first time by Bialynicki-Birula \cite{B-B}. 

Duality-invariant models of the form \eqref{models} have a long history. In 1999, Araki and Tanii \cite{AT} studied $\sU(N)$ duality-invariant theories\footnote{If the dimension of spacetime is even, $d=2n$, the maximal duality group for a system of $N$ gauge $(n-1)$-forms depends on the dimension of spacetime. The duality group is $\sU(N)$ if $n=2p$, 
and ${\sO}(N) \times {\sO}(N) $ if $n=2p+1$ \cite{AT}
(see, e.g., section 8 of \cite{KT2} for a review).
In the presence of scalars, the maximal duality group is $\sSp(2N, {\mathbb R})$ if $d = 4p$, and $\sSO(N,N)$
if $d=4p+2$.
The fact that the maximal duality group
depends on the dimension of space-time
was discussed in the mid-1980s \cite{Ta2,CFG} and also in 
the late 1990s \cite{CJLP1,CJLP2}.} 
of $N$ gauge $(n-1)$-forms with the filed strengths $F^I_{a(n)}$, $I=1,\dots,N$.
They restricted their discussion to Lagrangians 
\bea
L(F^I_{a(n)} ) = \cL (S^{IJ}, P^{IJ})~, \qquad 
S^{IJ} :=- \frac{1}{2n!} F^I\cdot F^J~, \qquad P^{IJ} :=- \frac{1}{2n!} F^I\cdot \widetilde{F}^J~,
\eea
and they used perturbation theory to demonstrate the existence of $\sU(N)$ duality-invariant theories possessing the weak-field limit, 
\bea
\cL = S^{II} + \cO(F^4)~.
\eea
No closed-form solution was given in \cite{AT}. Born-Infeld-like actions with $N$ field strengths $F^I_{a(n)}$ in $d=2n$ dimensions were studied in Zumino and collaborators in \cite{ABMZ}. In the $N=1$ case, their Born-Infeld $(n-1)$-form electrodynamics agreed with the one sketched by Gibbons and Rasheed \cite{GR1}.\footnote{In the $N=1$ case, 
the self-duality equations used in \cite{ABMZ} reduce to \eqref{self-duality3}.}  
The same model was also derived in 
\cite{Chruscinski:2000zm} using Hamiltonian techniques, and later in \cite{Buratti:2019cbm}. 
In fact, Buratti, Lechner and Melotti \cite{Buratti:2019cbm} also arrived at the equation \eqref{self-duality3} and the conclusion that every model for self-dual nonlinear electrodynamics admits an extension to higher dimensions.\footnote{This result is a by-product of similar observations for $\sU(1)$ duality-invariant higher-spin models in four dimensions \cite{Kuzenko:2026vir}.}

%%%%%%%%%%%%%%%%%%%%%%%%%%%%%%%%%%%%%%%%%%%%

\section{Examples of self-dual theories in $4p$ dimensions} 

Our discussion implies that every model for self-dual nonlinear electrodynamics in four dimensions 
admits an extension to self-dual $(n-1)$-form nonlinear electrodynamics in $ 2n=4p>4$ dimensions. 
In particular, associated with the Born-Infeld Lagrangian \cite{Born:1934gh} is the Born-Infeld $(n-1)$-form electrodynamics
\bea
L_{\rm BI} (F_{a(n)}) = T-\sqrt{T^{2} -2T S -P^2 }~,
\label{BI-HD}
\eea
with $T$ a duality-invariant parameter. This theory was introduced by several groups \cite{GR1, ABMZ, Chruscinski:2000zm, Buratti:2019cbm}.

Next, associated with the ModMax theory, which was discoved by Bandos, Lechner, Sorokin and Townsend
\cite{BLST} (and soon re-derived, in a simpler setting, by Kosyakov \cite{Kosyakov}), 
is the $\sU(1)$ duality-invariant $(n-1)$-form electrodynamics  \cite{Bandos:2020hgy}
\begin{align}\label{ModMax}
    {L}_{\rm MM}(F_{a(n)})
      = S \cosh\g  + \sqrt{S^{2}+P^{2}}  \sinh\g~,
\end{align}
which is conformal in $d=2n=4p$ dimensions. The original ModMax theory is a unique conformal and $\sU(1)$ duality-invariant theory in for dimensions. Its higher-dimensional extension \eqref{ModMax} is not unique. There exist more general self-dual and conformal models for $(n-1)$-form electrodynamics. 
This can be readily seen by making use of the auxiliary-field formulation of \cite{Kuzenko:2019nlm}
which is a generalisation of the Ivanov-Zupnik approach \cite{IZ0,IZ1,IZ2}.
In this formulation for $\sU(1)$ duality-invariant $(n-1)$-form nonlinear electrodynamics,  the Lagrangian 
$L(F, V)$  depends on an unconstrained  rank-$n$ 
antisymmetric tensor $V_{a_1 \dots a_n}$
and has the form
\bea
L(F,V) = \frac{1}{n!} \Big\{ \hf F \cdot F +  V \cdot V - 2 V \cdot F\Big\} 
+ L_{\rm int} (V) ~,
\label{first-order}
\eea
with $ V\cdot F:= V^{a_1 \dots a_n} F_{a_1 \dots a_n}$.
The self-duality equation \eqref{self-duality} is equivalent to the manifest $\sU(1)$ invariance of the self-interaction 
$ L_{\rm int} (V)= L_{\rm int} (V_+, V_-) $,
\bea
L_{\rm int} (\re^{\ri \vf}  V_+, \re^{-\ri \vf} V_-)  = L_{\rm int} (V_+, V_-) ~, \qquad 
\vf \in {\mathbb R}~.
\label{U(1)invariance}
\eea
The theory with Lagrangian \eqref{first-order} is conformal if the self-interaction is a homogeneous function of $V$ of degree $+2$, 
\bea
L_{\rm int} (\sigma V)=\sigma^2 L_{\rm int} (V)~.
\label{conformal}
\eea
We point out that the $\sU(1)$ duality-invariant theories of the form \eqref{models}
correspond to those auxiliary-field models \eqref{first-order} in which the self-interaction $L_{\rm int} (V) $
ls given by 
$ L_{\rm int} (V_+, V_-) = f (V_+ \cdot V_+ V_- \cdot V_-)$, with $f(x)$ a real function 
of one variable. 
The self-coupling corresponding to the $(n-1)$-form Mod-Max theory \eqref{ModMax} proves to be 
$L_{\rm int} (V) \propto \sqrt{V_+ \cdot V_+ V_- \cdot V_-}$, see \cite{K21} for the four-dimensional analysis. 
It is clear that in $d=2n=4p >4$ dimensions there exists more general solutions of the equations 
\eqref{U(1)invariance} and \eqref{conformal}.

Finally, we introduce a higher-dimensional generalisation of the algorithm proposed in \cite{Murcia:2025psi} to generate $\sU(1)$ duality-invariant models. Specifically, if the Lagrangian \eqref{models} is a solution of \eqref{self-duality3}, then the following model
\bea
 L_\g (F_{a(n)} ) := \cL (\O, P)~, \qquad \O:= S \cosh\g  + \sqrt{S^{2}+P^{2}}  \sinh\g
\eea
is also a  a solution of the self-duality equation \eqref{self-duality3}. 
This construction leads to new self-dual models for $(n-1)$-form electrodynamics.  
Applying this recipe to \eqref{BI-HD} leads to a higher-dimensional analog of the ModMaxBorn theory introduced in \cite{Bandos:2020hgy}.

%%%%%%%%%%%%%%%%%%%%%%%%%%%%%%%%%%%%%%%%%

\section{Energy-momentum flows}

In this paper we have demonstrated  that every model for self-dual nonlinear electrodynamics in four dimensions has a $\mathsf{U}(1)$ duality-invariant extension to $4p>4$ dimensions.\footnote{Many years ago it was shown \cite{KT1,KT2} that every model for self-dual nonlinear electrodynamics possesses an $\cN=1$ supersymmetric duality-invariant extension.}
Beyond four dimensions there exist more general self-dual theories, as follows from the known generating formulations for such theories 
\cite{Buratti:2019cbm, Kuzenko:2019nlm, Avetisyan:2022zza}.
The important point is that in $4p\equiv 2n>4$ dimensions there exist more general invariants of $F_{a(n)}$ (see \cite{Cederwall:2025ywy} for the technical details)
than the invariants $S$ and $P$ given in \eqref{models}.

A few years ago, a remarkable result was established \cite{Ferko:2023wyi}
for every model for  self-dual nonlinear electrodynamics ($n=2$).
Given a one-parameter family of $\sU(1)$ duality-invariant theories, $L(F;g)$, with $g$ a duality-invariant parameter, 
the Lagrangian obeys  $T\bar T$-like flow equation
\bea
\frac{\pa }{\pa g}  L  = \mathfrak{F} ( T_{ab} )~,
\label{flow}
\eea
for some function $\mathfrak{F} $ of   the energy-momentum tensor $T_{ab}$.
This theorem extends  several explicit examples considered earlier in the literature in the context of $T\bar T$ deformations \cite{Conti:2018jho, Babaei-Aghbolagh:2022uij, Ferko:2022iru}.
The quoted result of \cite{Ferko:2023wyi}
has been extended to nonlinear chiral theories in six dimensions \cite{Ferko:2024zth}. In ten dimensions, 
however, there is a flow in a chiral nonlinear theory which is not generated by the energy-momentum tensor \cite{Hutomo:2025dfx}, 
and a similar conclusion is expected in $4p$ dimensions. It is natural to wonder whether the relation \eqref{flow} holds for the family of self-dual theories considered in this paper. The answer appears to be negative in general.  
However, below we present a family of models for self-dual $(n-1)$-form electrodynamics in which the trace of the energy-momentum tensor determines the flow with respect to a duality-invariant deformation parameter. 

The energy-momentum tensor corresponding to $\cL(S,P)$, eq. \eqref{models},  is 
\bea
T_{ab} = \cL_S T_{ab}^{ (\text{free})} +\eta_{ab} \Big(\cL - S\cL_S -P\cL_P\Big) ~,
\label{EMT}
\eea
where 
\bea
T_{ab}^{ (\text{free})} = \frac{1}{(n-1)!} \Big( F_a \cdot F_b - \frac{1}{d} \eta_{ab} F\cdot F\Big) ~, \qquad 
F_a \cdot F_b := F_{a c_1\dots c_{n-1} } F_b{}^{c_1 \dots c_{n-1}} 
\eea
is the energy-momentum tensor of the free model \eqref{free}, see e.g. \cite{Kuzenko:2020zad}. 
Since \eqref{free} is conformal, the corresponding energy-momentum tensor is traceless, 
\bea
\eta^{ab} T_{ab}^{ (\text{free})} = 0~.
\eea
Making use of this property, for the energy-momentum tensor \eqref{EMT} we get
\bea
\Theta:= \eta^{ab}T_{ab} = d \Big(\cL - S\cL_S -P\cL_P\Big) ~.
\label{trace}
\eea
Conformal theories are characterised by the condition $\Theta =0$.

Now we should make two observations. Firstly, if $L(F_{a(n)} )$ is a solution of the self-duality equation
\eqref{self-duality}, then the following combination
 \bea
 L- \frac{1}{2 n!} F\cdot \widetilde{G} 
\eea 
is a duality-invariant observable \cite{AT}.\footnote{In the four-dimensional case ($n=2$) this result was derived by Gaillard and Zumino \cite{GZ1,GZ2,GZ3}.}  This means that it 
is invariant under the infinitesimal $\sSO(2)$ duality transformation 
\bea
 \d F = \l G ~, \qquad \d G =  - \l F~. 
\eea
In case $L(F_{a(n)} )$ is of the form \eqref{models}, one finds
 \bea
 L- \frac{1}{2 n!} F\cdot \widetilde{G} =\cL - S\cL_S -P\cL_P = \frac{1}{d} \Theta~.
 \eea

Secondly, we recall the observation given in \cite{KT2}, which is: if $L(F_{a(n)} )$ is a solution of the self-duality equation
\eqref{self-duality}, then 
\bea
L^{ (g) } (F_{a(n)}) := \frac{1}{g^2} L(g F_{a(n)})~, \qquad g \in {\mathbb R}^+ ~, 
\eea
is also a solution of the self-duality equation \eqref{self-duality}, with $g$ a duality-invariant parameter.\footnote{The Born-Infeld $(n-1)$-form electrodynamics \eqref{BI-HD} is an example of such a theory with $T=g^{-2}$.} 
According to  \cite{GZ2,GZ3,AT}, the operator $\pa  \cL^{ (g) } / \pa g$ is duality invariant. 
We now restrict our attention to the case of $L(F_{a(n)} )$ of the form  \eqref{models}. 
Then a direct calculation gives
\bea
\frac{ \pa  L^{(g)} }{ \pa g} = -\frac{2}{g} \Big\{ L^{(g)}- \frac{1}{2 n!} F\cdot \widetilde{G}^{ (g) } \Big\}~. 
\label{flow2}
\eea
The relations \eqref{trace} and \eqref{flow2} lead to the flow equation
\bea
\frac{\pa  L^{ (g) } }{ \pa g} = -\frac{2}{ g d} \Theta^{(g)}~,
\label{411}
\eea
which is a generalisation of the $d=4$ result derived in \cite{Ferko:2023wyi}.
In the case of the Born-Infeld $(n-1)$-form electrodynamics, eq. \eqref{BI-HD} was obtained in 
\cite{Babaei-Aghbolagh:2020kjg}. 

%%%%%%%%%%%%%%%%%%%%%%%%%%%%%%%%%%%%%%%%%%%%%%
%%%%%%%%%%%%%%%%%%%%%%%%%%%%%%%%%%%%%%%%%%%%%%

\section{Self-duality and quantum theory}

Models for $(n-1)$-form electrodynamics 
%The duality-invariant models for the gauge $(n-1)$-form $A_{n-1}$  
in $d=4p=2n>4$, including the self-dual theories studied above, 
are of some interest from the quantum mechanical point of view. Firstly, irrespective of duality invariance, any model of the form $L(F_n)$, with $F_n = \rd A_{n-1}$, is a reducible gauge theory,
 %of the $(n-2)$-th stage of reducibility, 
 following the
terminology of \cite{BV}.\footnote{Its stage of reducibility is $n-2$.
The field strength $F_n=\rd A_{n-1}$ is invariant under the gauge transformation 
$\delta A_{n-1} = \rd \zeta_{n-2} $. Here the variation $\d A_{n-1}$ remains unchanged if the gauge parameter is changed as 
$\zeta_{n-2} \to \zeta_{n-2} + \rd \zeta_{n-3}$, and so on. As a result, there emerge $n-2$ stages of reducibility.}   
Covariant quantisation of such theories cannot be carried out using the famous Faddeev-Popov approach \cite{Faddeev:1967fc} which is used for Yang-Mills theories. 
One should instead make use of alternative approaches\cite{Schwarz1,Schwarz2,Siegel, Thierry-Mieg:1980ihu, Obukhov,BK88}, which are applicable  when quantising reducible Abelian  gauge theories such as gauge $p$-forms, or the more powerful Batalin-Vilkovisy quantisation \cite{BV}.

In the case of self-dual $(n-1)$-form electrodynamics in $ 2n=4p>4$ dimensions, the quantisation aspects get entangled with dynamical ones. 
%Given a model for $\sU(1)$ duality-invariant $(n-1)$-form electrodynamics in $ 2n=4p>4$ dimensions, 
As a natural generalisation of the dilaton-axion coupling to self-dual nonlinear electrodynamics \cite{GR2,GZ2}, 
the $\sU(1)$ duality group can be enhanced to the non-compact group $\sSL(2,{\mathbb R})$ 
by coupling the gauge field $A_{n-1}$ to the dilaton $\vf$ and axion $\mathfrak a$ \cite{Tanii,AT}. In particular, associated withe the free $(n-1)$-form electrodynamics \eqref{free} is the interacting model\footnote{The second term in \eqref{dilaton-axion-coupling}
is identically zero in case that  $d=4p +2$.}
\bea
L (F,\tau, \bar \tau) = -\frac{1}{2 n!} \re^{-\vf} F^{a_1 \dots a_n}  F_{a_1 \dots a_n}  
+  \frac{1}{2n!} \mathfrak{a} \,\widetilde{F}^{a_1\dots a_n} F_{a_1\dots a_n}~, \qquad \tau=\mathfrak{a} +\ri \re^{-\vf}~.
\label{dilaton-axion-coupling}
\eea
The complex  field $\tau=\mathfrak{a} +\ri \re^{-\vf}$  is dimensionless and parametrises the Hermitian symmetric space 
 $\sSL(2,{\mathbb R})/ \sSO(2)$ realised as the Poincar\'e upper half-plane.
 The duality group $\sSL(2,{\mathbb R})$ acts on $\tau$ by fractional linear transformations
\bea
\t' = \frac{a\t +b}{c\t+d}~, \qquad
\left( \begin{array}{cc} a& b \\ c & d \end{array} \right) \in
\sSL(2, {\mathbb R})~.
\eea
At the classical level, the theory \eqref{dilaton-axion-coupling} is conformal and possesses $\sSL(2,{\mathbb R})$ duality invariance.
Both symmetries prove to be anomalous in the quantum theory. However, if one integrates out the potential $A_{n-1}$ in the path integral, it is known that the logarithmically divergent part of the one-loop effective action (the so-called induced action) should be a local functional of $\tau $ and $\bar \tau$ that is conformal and $\sSL(2,{\mathbb R})$ invariant. 
This induced action has not yet been computed for $n>2$, and its derivation is an interesting open problem. 
In addition, to the best of our knowledge, covariant quantisation of the gauge theory \eqref{dilaton-axion-coupling} has not been carried out for $n>2$.

In four dimensions ($n=2)$, the induced action was computed  by Osborn \cite{Osborn}, and re-derived later in \cite{BPT} in the framework of induced ${\cal N}=4$ conformal supergravity. In curved space, its general form is 
\bea
{L}^{(n=2)}_{\rm induced} &=&  \frac{1}{2( {\rm Im}\, \t)^2} \Big[ \cD^2 \t \cD^2 \bar \t 
- 2 (R^{ab}- \frac 13 \eta^{ab} R) \nabla_a \t \nabla_b \bar \t \Big] \non \\
&&+ \frac{1}{12 ( {\rm Im}\, \t)^4} 
\Big[ \a \nabla^a \t \nabla_a \t \nabla^b \bar \t  \nabla_b \bar \t 
+\b \nabla^a \t \nabla_a \bar  \t \nabla^b  \t  \nabla_b \bar \t 
\Big] 
%\label{1.2}
\eea
where $\nabla_a$ is the torsion-free Lorentz-covariant derivative,\footnote{As usual, 
$\nabla_a = e_a{}^m (x)  \pa_m +\hf \omega_a{}^{bc} (x)M_{bc}$, where $e_a{}^m$ is the inverse vielbein, 
$ \omega_a{}^{bc} $ the Lorentz connection, 
and $M_{bc} $ the Lorentz generator. The algebra of covariant derivatives is $[\nabla_a, \nabla_b] = \hf R_{ab}{}^{cd}M_{cd}$,
with $R_{ab}{}^{cd}$ the Riemann tensor.}
\bea
\cD^2 \t := 
%\Big(\nabla^a + \frac{\ri}{{\rm Im}\, \t} \nabla^a \t  \Big)\nabla_a \t  \equiv 
\hat{\nabla}{}^a \nabla_a \tau~,
\qquad \hat{\nabla}_a  v^\tau=\Big( \nabla_a 
%\nabla_a \t 
+ \frac{\ri}{{\rm Im}\, \t} \nabla_a \Big) v^\tau ~, \quad 
\hat{\nabla}_a  v^{\bar \tau}=\Big( \nabla_a 
%\nabla_a \t 
- \frac{\ri}{{\rm Im}\, \t} \nabla_a \Big) v^{\bar \tau} ~,
\eea
and $\a$ and $\b$ are numerical parameters. Here $\hat{\nabla}_a$ is the standard $\sigma$-model covariant derivative 
corresponding to the K\"ahler metric on the Poincar\'e upper half-plane
\bea 
\rd s^2 = \frac{1}{ ({\rm Im}\, \tau)^2} \rd \tau \rd \bar \tau = 2 g_{\tau \bar \tau} \rd \tau \rd \bar \tau~,
\label{metric}
\eea
with the Christoffel symbols being 
 $\Gamma^\tau{}_{\tau\tau} = {\rm i} ( {\rm Im}\,\tau)^{-1}$ and   
 $\Gamma^{\bar \tau}{}_{\bar \tau \bar \tau} = -{\rm i} ( {\rm Im} \,  \tau)^{-1}$. 
The functional $ \int \rd^4 x \, e\, {L}^{(n=2)}_{\rm induced} $ is manifestly 
 $\sSL(2,{\mathbb R})$  invariant and also proves to be Weyl invariant.
The Weyl invariance follows from the fact that the Fradkin-Tseytlin (FT)
operator \cite{FT1982}
\bea
\Delta_4 = (\nabla^a \nabla_a)^2 + 2 \nabla^a \big(
	 {R}_{ab} \,\nabla^b 
	- \tfrac{1}{3} {R} \,\nabla_a
	\big)
\label{1.5}
\eea
is conformal.\footnote{This operator was  re-discovered 
by Paneitz in 1983 \cite{Paneitz} and Riegert in 1984 \cite{Riegert}.}

In the $d=4$ case, the simple model \eqref{dilaton-axion-coupling} 
may be generalised to a system of $n$ Abelian vector fields coupled to $\frac 12 n(n+1)$ complex scalars parametrising the Hermitian symmetric space
$\mathsf{Sp}(2n, {\mathbb R})/ \mathsf{U}(n)$. Such a theory is conformal invariant and possesses the maximal non-compact duality group $\mathsf{Sp}(2n, {\mathbb R})$. The corresponding induced action, obtained by integrating out the vector fields, was computed in 
\cite{Grasso:2023hmv}. The induced action, which determines the logarithmically divergent part of the one-loop effective action, is conformal  and $\mathsf{Sp}(2n, {\mathbb R})$ invariant.

Beyond four dimensions, $n>2$, the induced action corresponding to \eqref{dilaton-axion-coupling}
should also be  $\sSL(2,{\mathbb R})$  and Weyl invariant. This means that the effective Lagrangian ${L}^{(n>2)}_{\rm induced} $ describes a higher-derivative nonlinear $\sigma$-model with leading part
\bea
{L}^{(n>2)}_{\rm induced} &=&  \frac{1}{2( {\rm Im}\, \t)^2}  \Big[\big(\hat{\nabla}{}^2\big)^{n/2-1}\cD^2 \t  \Big]
\big(\hat{\nabla}{}^2\big)^{n/2-1}\cD^2 \bar \t +\dots
\eea
The existence of such a Lagrangian is guaranteed by the existence of a conformal differential operator in $d=2n$ dimensions of the form \cite{GJMS} (see also \cite{Gover:2002ay, Manvelyan:2007tk, Juhl:2011ua} and references therein)
\bea
\Delta_{d}=(\nabla^a \nabla_a)^{d/2} +\dots 
\eea
where the ellipsis denotes terms of lower order in the covariant derivatives.\footnote{In six dimensions, the operator $\Delta_6$
was constructed in \cite{Branson,Wunsch}.}
\\
%%%%%%%%%%%%%%%%%%%%%%%%%%%%%%%%%%%
%%%%%%%%%%%%%%%%%%%%%%%%%%%%%%%%%%%

\noindent
{\bf Acknowledgements:}\\ 
I am grateful to Jessica Hutomo for bringing Ref. \cite{Babaei-Aghbolagh:2020kjg} to my attention, and to Dmitri Sorokin for useful discussions and for attracting my renewed attention to \cite{Buratti:2019cbm}.
This work is supported in part by the Australian Research Council, project DP230101629.

\begin{footnotesize}

\end{footnotesize}

%%%%%%%%%%%%%%%%%%%%%%%%%%%%%%%%%%%%
%%%%%%%%%%%%%%%%%%%%%%%%%%%%%%%%%%%%

\end{document}